\documentstyle[epsfig,twocolumn]{esapub}

\def\HI{\hbox{H~$\scriptstyle\rm I\ $}}
\def\HII{\hbox{H~$\scriptstyle\rm II\ $}}

\def\nH{{\rm H}}
\def\nHII{{\rm HII}}

\def\HeII{\hbox{He~$\scriptstyle\rm II\ $}}

\def\HeIII{\hbox{He~$\scriptstyle\rm III\ $}}

\def\kms{\,{\rm km\,s^{-1}}}
\def\kmsmpc{\,{\rm km\,s^{-1}\,Mpc^{-1}}}

\def\emunits{\,{\rm ergs\,s^{-1}\,Hz^{-1}\,Mpc^{-3}}}
\def\ndotunits{\,{\rm s^{-1}\,Mpc^{-3}}}

\def\msun{\,{\rm M_\odot}}

\def\sfrd{\,{\rm M_\odot\,yr^{-1}\,Mpc^{-3}}}
\def\sfr{\,{\rm M_\odot\,yr^{-1}}}

\def\Lya{Ly$\alpha\ $}
\def\etal{{et al.\ }}
\def\AB{{\rm AB}}
\def\spose#1{\hbox to 0pt{#1\hss}}
\def\lta{\mathrel{\spose{\lower 3pt\hbox{$\mathchar"218$}}
     \raise 2.0pt\hbox{$\mathchar"13C$}}}
\def\gta{\mathrel{\spose{\lower 3pt\hbox{$\mathchar"218$}}
     \raise 2.0pt\hbox{$\mathchar"13E$}}}

\def\micron{\,$\mu${\rm m}}

\begin{document}
\setlength{\parindent}{0pt}
\setlength{\parskip}{ 10pt plus 1pt minus 1pt}
\setlength{\hoffset}{-1.5truecm}
\setlength{\voffset}{-2.0truecm}
\setlength{\textwidth}{17.1truecm }
\setlength{\columnsep}{1truecm }
\setlength{\columnseprule}{0pt}
\setlength{\headheight}{12pt}
\setlength{\headsep}{20pt}
\pagestyle{esapubheadings}

\title{\bf PROBING THE HIGH REDSHIFT UNIVERSE WITH THE NGST}
\author{{\bf Piero Madau} \vspace{2mm} \\
Space Telescope Science Institute, 3700 San Martin 
Drive, Baltimore, MD 21218, USA}

\maketitle
\abstract{
Through a combination of deep wide-field imaging and
multi-object spectroscopy, the Next Generation Space Telescope will be able 
to chart with unprecedented accuracy the evolution of cosmic structures 
after the `dark ages' ($z\lta 5$), when galaxies are thought to assemble and 
form the bulk of their stars. In this talk I will discuss how NGST observations
might determine the history of the stellar birthrate in the universe and  
reveal the character of cosmological ionizing sources at high redshifts.}

\section{INTRODUCTION}

The remarkable progress in our understanding of faint galaxy data made possible
by the combination of HST deep imaging (Williams \etal 1996) and 
ground-based spectroscopy (Lilly \etal 1996; Ellis \etal 1996; Steidel \etal 
1996b), has recently permitted to shed
some light on the evolution of the stellar birthrate in the universe, to 
identify the epoch $1\lta z\lta 2$ where most of the optical extragalactic 
background light was produced, and to set important contraints on galaxy 
formation scenarios (e.g., Madau \etal 1998c; Steidel \etal 1998). While one 
of the
topical questions concerning our understanding of the emission history of the 
universe is the nature and redshift distribution of the recently discovered 
population of sub-mm sources (Hughes \etal 1998; Barger \etal 1998; Lilly, this
volume), one could also imagine the existence of a large population of faint 
galaxies still undetected at high-$z$, as the color-selected ground-based and 
{\it Hubble Deep Field} (HDF) samples include only the brightest star-forming 
young 
objects. In hierarchical clustering scenarios, a population of high-$z$ dwarfs 
(i.e. an early generation of stars in dark matter halos with circular velocities
$v_{\rm circ} \approx 50\,\kms$) is actually expected to be one of the main 
source of UV photons and heavy elements at early epochs (e.g. Miralda-Escud\'e 
\& Rees 1998; Loeb, this volume). 

While it is hard to predict in detail what the scientific focus of NGST will
be ten years from now in a rapidly evolving area like physical cosmology, I 
believe it is a useful exercise to describe a couple of illustrative science 
programs which well highlight  
the strategic capabilities of NGST in providing a comprehensive 
inventory of the buildup of metals, stars, and light in the universe: {\it (a)}
the determination of the rate of Type Ia and Type II supernova (SN) explosions 
as a function of cosmic time from intermediate to high redshifts; and {\it (b)}
the study of candidate sources of photoionization at early times.
Throughout this paper I will adopt an Einstein-de Sitter universe ($q_0=0.5$) 
with $H_0=50h_{50}\, \kmsmpc$. 

\section{SUPERNOVAE}

The evolution of the SN rate with redshift contains unique
information on the star formation history of the universe, the initial mass
function (IMF) of stars, and the nature of the binary companion in Type Ia
events. All are essential ingredients for understanding galaxy formation,
cosmic chemical evolution, and the mechanisms which determined the efficiency
of the conversion of gas into stars in galaxies at various epochs. While the
frequency of ``core-collapse supernovae'', SN~II and possibly SN~Ib/c, which
have short-lived progenitors, is
essentially related, for a given IMF, to the instantaneous stellar birthrate of
massive stars, Type Ia SNe -- which are believed to result from the
thermonuclear disruption of C-O white dwarfs in binary systems -- follow a
slower evolutionary clock, and can then be used as a probe of the past history
of star formation in galaxies (e.g., Yungelson \& Livio 1998). 
The recent detection of Type Ia SNe at cosmological distances (Garnavich 
\etal 1998; Perlmutter \etal 1998) allows for the first time a 
detailed comparison at $z\lta 1$ between the SN rates
self-consistently predicted by stellar evolution models that reproduce the
optical spectrophotometric properties of field galaxies,
and the observed values. 
\begin{figure*}
\centerline{\epsfig{file=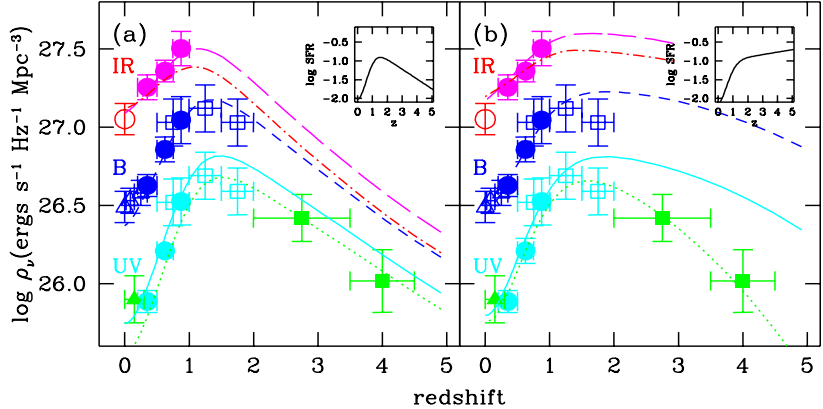,height=4.3in,width=4.5in}}
\vspace{-4.5cm}
\caption{\em Evolution of the observed comoving luminosity density at 
rest-frame 
wavelengths of 0.15 ({\it dotted line}), 0.28 ({\it solid line}), 0.44 
({\it short-dashed line}), 1.0 ({\it long-dashed line}), and 2.2 ({\it 
dot-dashed line}) \micron. The data points with error bars are taken from 
Lilly \etal (1996) ({\it filled dots}), Connolly \etal (1997) ({\it empty 
squares}), Madau \etal (1996) ({\it filled squares}), Ellis 
\etal (1996) ({\it empty triangles}), Treyer \etal (1997) ({\it filled 
triangle}), and Gardner \etal (1997) ({\it empty dot}). ({\it a}) This model 
assumes a Salpeter IMF, SMC-type dust in a foreground screen, and a 
universal $A_{1500}=0.8$ mag. ({\it b}) This model -- designed to mimick a
monolithic collapse scenario -- assumes a Salpeter IMF and a dust opacity
which increases rapidly with redshift, $A_{1500}=0.09(1+z)^{2.2}$ mag. 
\label{fig1}}
\end{figure*}
Accurate measurements at all redshifts  of the frequencies
of Type II(+Ib/c) and Ia SNe could be used as an independent test for the star
formation and heavy element enrichment history of the universe,  and 
significantly improve our understanding of the intrinsic nature and age 
of the populations involved in the SN explosions (Sadat \etal 1998; Madau
\etal 1998a). A determination of the amount 
of star formation at early epochs may be of crucial importance, as the two 
competing scenarios for galaxy formation, monolithic collapse -- where 
spheroidal systems formed early and rapidly, experiencing a bright starburst 
phase at high-$z$ (Eggen \etal 1962; Tinsley \& Gunn 
1976) -- and hierarchical clustering -- where ellipticals form continuosly by 
the merger of disk/bulge systems (White \& Frenk 1991; Kauffmann \etal 1993) 
and most galaxies never experience star formation rates in excess of a few 
solar masses per year (Baugh
\etal 1998) -- appear to make rather different predictions in this regard. By 
detecting Type II SNe at high-$z$, NGST 
should provide an important test for distinguishing between different 
scenarios of galaxy formation. 

\subsection{Cosmic Star Formation History}

The emission history of field galaxies at ultraviolet, optical, and 
near-infrared wavelengths can be modeled by tracing the evolution with cosmic 
time of their luminosity density, 
\begin{equation}
\rho_\nu(z)=\int_0^\infty L_\nu y(L_\nu,z)dL_\nu=\Gamma(2+\alpha)y_*L_*,
\end{equation}
where $y(L_\nu,z)$ is the best-fit Schechter luminosity function in each
redshift bin. The integrated light radiated per unit volume from the entire
galaxy population is an average over cosmic time of the stochastic, possibly
short-lived star formation episodes of individual galaxies, and follows a
relatively simple dependence on redshift.  A stellar evolution model, defined 
by a time-dependent star formation rate per 
unit volume, $\psi(t)$, a universal IMF, $\phi(m)$, and some amount of dust 
reddening, can actually reproduce the optical data reasonably well (Madau
\etal 1998c). In such a
system, the luminosity density at time $t$ is given by the convolution integral
\begin{equation} 
\rho_\nu(t)=p_{\rm esc}\int^t_0 l_\nu(t')\psi(t-t')dt', 
\end{equation} 
where $l_\nu(t')$ is the specific luminosity radiated per unit initial mass
by a generation of stars with age $t'$, and $p_{\rm esc}$ is a
time-independent term equal to the fraction of emitted photons which are not
absorbed by dust. 
The function $\psi(t)$ is derived from the observed UV luminosity density, and
is then used as input to the population synthesis code of Bruzual \& Charlot 
(1998). 
\begin{figure*}
\centerline{\epsfig{file=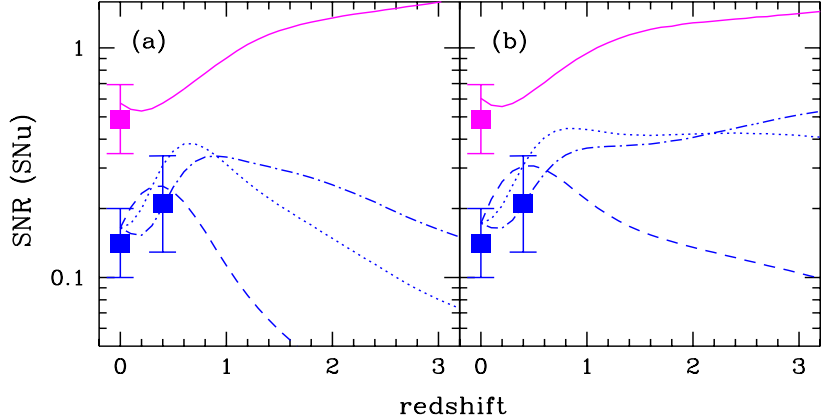,height=4.3in,width=4.5in}}
\vspace{-4.5cm}
\caption{\em Predicted Type Ia and II(+Ib/c) rest-frame frequencies as 
a function of redshift. The rates are normalized to the {\it emitted} blue 
luminosity 
density. {\it Solid line}: SN II rate.	{\it Dashed-dotted line}: SN Ia 
rate with $\tau=0.3$ Gyr. {\it Dotted line}: SN Ia rate with $\tau=1$ Gyr. {\it Dashed line}: SN Ia rate with $\tau=3$ Gyr. The data points 
with error bars have been derived from the measurements of Cappellaro 
\etal (1997), Tammann \etal (1994), Evans \etal (1989), and Pain \etal (1997), and have been weighted according to 
the local blue luminosity function by spectral type of Heyl \etal (1997). ({\it a}) Model predictions for the merging scenario of Figure 1{\it a}.  ({\it b}) 
Same for the monolithic collapse scenario of Figure 1{\it b}.
\label{fig2}}
\end{figure*}
Figure 1{\it a} shows the model predictions for the evolution of $\rho_\nu$ for
a Salpeter function, $A_{1500}=0.8$ mag with SMC-type dust, and a star 
formation history which traces the rise,
peak, and sharp drop of the UV emissivity. For simplicity, the metallicity was fixed to solar values and the IMF truncated at 0.1
and 125 $\msun$. The data points show the observed luminosity density in six
broad passbands centered around 0.15, 0.20, 0.28, 0.44, 1.0, and 2.2 \micron. 
The model is able to account for the entire background light recorded in 
the galaxy counts down to the very faint magnitude levels probed by the 
HDF, and produces visible mass-to-light ratios at the present epoch
which are consistent with the values observed in nearby galaxies of various
morphological types. The bulk ($\gta 60\%$ by mass) of the stars present today formed relatively recently ($z\lta
1.5$), consistently with the expectations from a broad class of hierarchical
clustering cosmologies (Baugh \etal 1998), and in good agreement with the low
level of metal enrichment observed at high redshifts in the damped \Lya 
systems (Pettini \etal 1997). 

One of the  biggest uncertainties in our understanding of the evolution of
luminous matter in the universe is represented by the poorly constrained
amount of starlight that was absorbed by dust and reradiated in the far-IR
at early epochs. Figure 1{\it b} shows the model predictions for a monolithic collapse 
scenario, where half of the present-day stars were formed at $z>2.5$ and
were enshrouded by an increasing amount of dust, 
$A_{1500}=0.09(1+z)^{2.2}$ mag. This results in a correction to the rate of 
star formation by a factor of $\approx 5$ at $z=3$ and $\approx 15$ at $z=4$.
The two models in Figure 1 produce a significant fraction ($\sim 50\%$) of the 
IR background detected by COBE (Dwek \etal 1998; Fixsen \etal 1998).

\subsection{Type Ia and II(+Ib/c) Supernova Rates}

Single stars with mass $>8\msun$ evolve rapidly ($\lta 50\,$ Myr) through all
phases of central nuclear burning, ending their life as Type II SNe with
different characteristics depending on the progenitor mass. For a Salpeter IMF
the core-collapse
supernova rate can be related to the stellar birthrate according to 
\begin{equation}
{\rm SNR}_{\rm II}(t)=\psi(t){\int_8^{125} dm \phi(m)\over \int_{0.1}^{125} 
dm m \phi(m)}.
 \end{equation} 

The specific evolutionary history leading to a Type Ia event remains instead 
an unsettled question. SN Ia are believed to result from the explosion of C-O
white dwarfs (WDs) triggered by the accretion of material from a companion, the
nature of which is still unknown (see Ruiz-Lapuente \etal 1997 for a recent
review). In a {\it double degenerate} (DD) system, for example, such elusive
companion is another WD: the exploding WD reaches the Chandrasekhar limit and
carbon ignition occurs at its center. In the {\it single
degenerate} (SD) model instead, the companion is a nondegenerate, evolved star
that fills its Roche lobe and pours hydrogen or helium onto the WD 
(Iben \& Tutukov 1984). While in the latter the clock for the
explosion is set by the lifetime of the primary star, and, e.g., by how long it
takes to the companion to evolve and fill its Roche lobe, in the former it is
controlled by the lifetime of the primary star and by the time it takes to shorten the separation of the two WDs as a
result of gravitational wave radiation. The evolution of the rate depends then,
among other things, on the unknown mass distribution of the secondary binary
components in the SD model, or on the distribution of the initial separations
of the two WDs in the DD model. 

To shed light into the identification issue and, in particular, on the 
clock-mechanism for the explosion of Type Ia's, it is useful to parametrize 
the rate of Type Ia's in terms of 
a characteristic explosion timescale, $\tau$ -- which defines an explosion
probability per WD assumed to be {\it independent} of time -- and an explosion
efficiency, $\eta$. The former accounts for the time it takes in the various
models to go from a {\it newly born} (primary) WD to the SN explosion itself: a
spread of ``delay'' times results from the combination of a variety of initial
conditions, such as the mass ratio of the binary system, the distribution of
initial separations, the influence of metallicity on the mass transfer rate and
accretion efficiency, etc. The latter simply accounts for the fraction of stars 
in binary systems that, because of unfavorable
initial conditions, will never undergo a SN Ia explosion. Possible progenitors 
are all systems in which the primary star has an {\it initial}
mass higher than $m_{\rm min}=3\,M_\odot$ (final mass $\ge 0.72\,M_\odot$,
Weidemann 1987) and lower than $m_{\rm max}=8\,M_\odot$: stars less massive
than $3\,M_\odot$ will not produce a catastrophic event even if the companion
has comparable mass, while stars more massive than $8\,M_\odot$ will undergo
core collapse, generating a Type II explosion. 
\begin{figure*}
\centerline{\epsfig{file=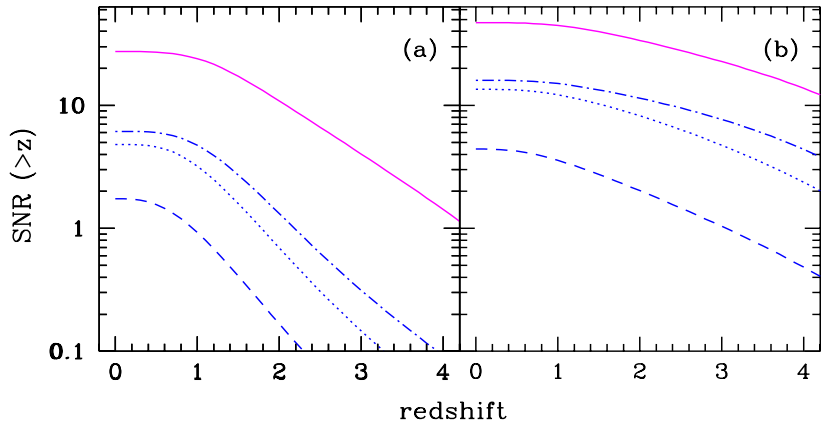,height=4.3in,width=4.5in}}
\vspace{-4.5cm}
\caption{\em Predicted cumulative number of Type Ia and II(+Ib/c) SNe above a
given redshift $z$ in a $4'\times 4'$ NGST field.  {\it Solid line}: Type II's.
{\it Dashed-dotted line}: Type Ia's with $\tau=0.3$ Gyr. {\it Dotted 
line}: Type Ia's with $\tau=1$ Gyr. {\it Dashed line}: Type Ia's with 
$\tau=3$ Gyr.  The effect of dust extinction on the detectability of SNe has
not been included in the models. ({\it a}) Model predictions for the 
merging scenario of Figure 1{\it a}.  ({\it b}) Same for the monolithic 
collapse scenario of Figure 1{\it b}.
\label{fig3}}
\end{figure*}
With these assumptions the rate of Type Ia events at any one time will be
given by the sum of the explosions of all the binary WDs produced in the
past that have not had the time to explode yet, i.e. 
\begin{equation}
{\rm SNR}_{\rm Ia}(t)={\eta \int^t_0 \psi(t')dt'\int_{m_c} 
^{m_{\rm max}} \exp(-T)\phi(m)dm \over \tau \int\phi(m)dm},
\end{equation}
where $m_c\equiv{\rm max}[m_{\rm min},m(t')]$, $m(t')=(10\,{\rm Gyr}/t')^{0.4}$ 
is the minimum mass of a star that 
reaches the WD phase at time $t'$, $t_m=10\,{\rm Gyr}/m^{2.5}$ is the 
standard lifetime of a star of mass $m$ (all stellar masses are expressed 
in solar units), and $T=(t-t'-t_m)/\tau$. 
For a fixed initial mass $m$, the frequency of Type Ia events
peaks at an epoch that reflects an ``effective'' delay $\Delta t\approx 
\tau+t_m$
from stellar birth. A prompter (smaller $\tau$) explosion results in a 
higher SN Ia rate at early epochs. 

In Figure 2 the predicted Type Ia and II(+Ib/c) rest-frame frequencies are
shown as a function of redshift. Expressed in SNu (one SNu corresponding
to 1 SN per 100 years per $10^{10} L_{B\odot}$), the Type II rate
is basically proportional to the ratio between the UV and blue galaxy
luminosity densities, and is therefore independent of cosmology. Unlike the SN
frequency per unit volume, which will trace the evolution of the stellar 
birthrate, the  frequency of Type II events per unit blue luminosity
is a monotonic increasing function of redshift, and depends only weakly on the 
assumed star formation history. The Type Ia rates plotted in the figure
assume characteristic ``delay'' timescales after the collapse of the primary 
star to a WD equal to $\tau=0.3, 1$ and 3 Gyr, which virtually encompass 
all relevant possibilities. The SN Ia explosion efficiency was left as an
adjustable parameter to reproduce the observed ratio of SN II to SN Ia
explosion rates in the local Universe (${\rm SNR}_{\rm II}/{\rm SNR}_{\rm Ia}
\approx 3.5$), $5\%<\eta<10\%$ for the adopted models. 
It appears that observational determinations of the SN~Ia rate at $z\sim1$ can 
unambiguously identify the appropriate delay time.  In particular, we 
estimate that measuring the frequency of Type Ia events at both $z\sim0.5$ 
and $z\sim 1$ with an error of 20\% or lower would allow one to determine 
this timescale to within about 30\%.  This kind of observations are by no 
means prohibitive, and these goals could be achieved within a couple of years. 
In fact, ongoing searches for high-$z$ SNe are currently able to discover and study about a 
dozen new events per observing session in the redshift range 0.4--1.0, and the 
observations are carried out at a rate of about four sessions a year. Since 
determining the frequency of SN~Ia with a 20\% uncertainty requires 
statistics on more than 25 objects per redshft bin, it is clear that, barring 
systematic biases, those rates will soon be know with high accuracy. Also note 
how, relative to the merging scenario, the monolithic collapse model predicts 
Type Ia rates (in SNU) that are, in the $\tau=0.3\,$ Gyr case, a factor of 1.6 
and 4.9 higher at $z=2$ and 4, respectively, with even larger factors found 
in the case of longer delays.  Therefore, once such timescale is calibrated
through the observed ratio SNR$_{\rm Ia}(z=0)/$SNR$_{\rm Ia}(z=1)$, 
one should be able to constrain the star formation 
history of the early universe by comparing the predicted SN~Ia rate at $z>2$
with the observations. 

The main results of this exercise can be summarized as follows:

\begin{itemize}

\item At the present epoch, the predicted Type II(+Ib/c) frequency matches 
remarkably well the observed local value. Note, however, that rates obtained 
from traditional distant (beyond 4 Mpc) sample 
might need to be increased by a factor of 1.5--2 because of severe selection
effects against Type II's fainter than $M_V=-16$ (Woltjer 1997).

\item In the interval $0\lta z\lta 1$, the predicted rate of SN Ia  
is a sensitive function of the characteristic delay timescale between 
the collapse of the primary star to a WD and the SN event. Accurate 
measurements of SN rates in this redshift range will improve 
our understanding of the nature of SN~Ia progenitors and the physics of 
the explosions. Ongoing searches and studies of distant SNe should soon 
provide these rates, allowing a universal calibration of the Type Ia phenomenon.

\item While Type Ia rates at $1\lta z\lta 2$ will offer valuable information 
on the star formation history of the universe at earlier epoch, 
the full picture will only be obtained with statistics on Type Ia and
II SNe at redshifts $2<z<4$ or higher. At these epochs, the  
detection of Type II events must await NGST. A SN II has a typical peak
magnitude $M_B\approx -17$: placed at $z=3$, such an
explosion would give rise to an observed flux of 15 nJy (assuming a
flat cosmology with $q_0=0.5$ and $H_0=50\,h_{50}\kmsmpc$) at 1.8 \micron.
At this wavelength,
the imaging sensitivity of an 8m NGST is 1 nJy ($10^4$ s exposure and
$10\sigma$ detection threshold), while the moderate resolution ($\lambda/\Delta
\lambda=1000$) spectroscopic limit is about 50 times higher ($10^5$ s exposure
per resolution element and $10\sigma$ detection threshold) (Stockman \etal
1998). The several weeks period of peak rest-frame blue luminosity would be
stretched by a factor of $(1+z)$ to few months. Figure 3 shows the cumulative
number of Type II events expected per year per $4'\times 4'$ field. Depending
on the history of star formation at high redshifts, NGST should detect
between 7 (in the merging model) and 15 (in the monolithic collapse scenario)
Type II SNe per field per year in the interval $2<z<4$. The
possibility of detecting Type II SNe at $z\gta5$ from an early population of
galaxies has been investigated by Miralda-Escud\'e \& Rees (1997). By assuming
these are responsible for the generation of all the metals observed in the
\Lya forest at high redshifts, a high baryon density
($\Omega_bh_{50}^2=0.1$), and an average metallicity of $0.01Z_\odot$,
Miralda-Escud\'e \& Rees estimate NGST should observe about 16 SN II per
field per year with $z\gta 5$. Note, however, that a metallicity smaller by a
factor $\sim 10$ compared to the value adopted by these authors has been
recently derived by Songaila (1997). For comparison, the models discussed above
predict between 1 and 10 Type II SNe per field per year with $z\gta 4$. 

\end{itemize}

\section{WHAT KEEPS THE UNIVERSE IONIZED AT $Z=5$?}

The existence of a filamentary, low-density intergalactic medium (IGM), which
contains the bulk of the hydrogen and helium in the universe, is predicted as a
product of primordial nucleosynthesis and of hierarchical 
models of gravitational instability with ``cold dark matter'' (CDM) (Cen \etal 
1994; Zhang \etal 1995; Hernquist \etal 1996). The application of the 
Gunn-Peterson constraint on the amount of smoothly distributed neutral 
material along the line of sight to distant objects requires the hydrogen 
component of the diffuse IGM to have been highly ionized by $z\approx 5$ 
(Schneider \etal 1991), and the helium component by $z\approx 2.5$ (Davidsen
\etal 1996).  The plethora of discrete absorption systems which give origin 
to the \Lya forest in the spectra of background quasars are also inferred 
to be strongly photoionized: we know from QSO absorption studies, in fact, that 
neutral hydrogen accounts for only a small fraction, $\sim 10\%$, of the 
nucleosynthetic baryons at early epochs (Lanzetta \etal 1995). It thus appears 
that substantial sources of ultraviolet photons were present at $z>5$, perhaps 
low-luminosity quasars (Haiman \& Loeb 1998) or a first generation of stars in 
virialized dark matter halos with $T_{\rm vir}\sim 10^4-10^{5} \,$K (Couchman
\& Rees 1986; Ostriker \& Gnedin 1996; Haiman \& Loeb 1997; Miralda-Escud\'e 
\& Rees 1998): early star formation provides a possible explanation for the 
widespread existence of heavy elements in the IGM (Cowie \etal 1995).
More in general, establishing the character of
cosmological ionizing sources is an efficient way to constrain competing
models for structure formation in the universe, and to study the collapse and
cooling of small mass objects at early epochs. 

\subsection{Quasars}

A decline in the space density of bright quasars at redshifts
beyond $\sim 3$ was first suggested by Osmer (1982), and has been since then
the subject of a long-standing debate. In recent years, several optical surveys
have consistently provided new evidence for a turnover in the QSO counts 
(Hartwick \& Schade 1990, HS; Warren \etal 1994, WHO; Schmidt \etal 1995, SSG; 
Kennefick \etal 1995, KDC).  The interpretation of the drop-off observed in 
optically selected samples is
equivocal, however, because of the possible bias introduced by dust obscuration
arising from intervening systems (Ostriker \& Heisler 1984). Radio emission, 
on the other hand, is unaffected by dust, and it has recently been shown 
(Shaver \etal 1996) that the space density of radio-loud quasars also 
decreases strongly for $z>3$ (Fig. 4), 
demonstrating that the turnover is indeed real and that dust along the line 
of sight has a minimal effect on optically-selected QSOs. 
\begin{figure}
\centerline{\epsfig{file=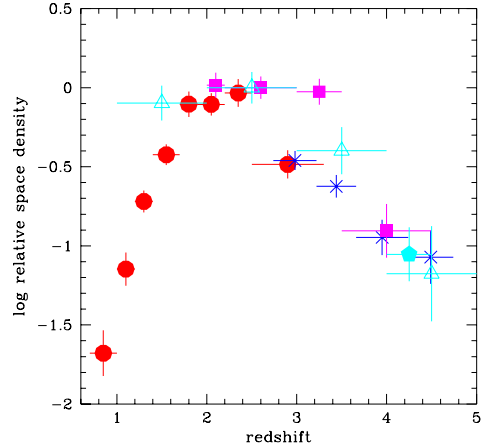,height=7.5cm,width=7.5cm}}
\caption{\em Comoving space density of bright QSOs as a function of redshift.
The data points with error bars are
taken from HS {\it (filled dots)}, WHO {\it (filled 
squares)}, SSG {\it (crosses)}, and KDC {\it (filled
pentagon)}. The points have been normalized to the $z=2.5$ space density of
quasars with $M_B<-26$ ($M_B<-27$ in the case of KDC). The {\it empty
triangles} show the space density (normalized to the peak) of the Parkes
flat-spectrum radio-loud quasars with $P>7.2\times 10^{26}\,$ W Hz$^{-1}$
sr$^{-1}$ (Hook \etal 1998).
\label{fig4}}
\end{figure}
The QSO emission rate of hydrogen ionizing photons per unit comoving volume,
$\dot{\cal N}$, has been recently recalculated by Madau \etal (1998b), and 
is shown in Figure 
5. It is important to notice that the procedure adopted to derive this 
quantity implies a large correction for incompleteness at high-$z$. With a fit 
to the quasar luminosity function (LF)  which goes as $\phi(L)\propto 
L^{-1.64}$ at the faint end (Pei 1995), the
contribution to the emissivity converges rather slowly, as $L^{0.36}$. At
$z=4$, for example, the blue magnitude at the break of the LF is $M_*\approx 
-25.4$, comparable or slightly fainter than the limits of current high-$z$  
QSO surveys. A large fraction, about 90\% at $z=4$ and even higher at 
earlier epochs, of the ionizing emissivity in our model is therefore 
produced by quasars that have not been actually observed, and are
assumed to be present based on an extrapolation from lower redshifts. The 
value of $\dot{\cal N}$ obtained by including the contribution from 
{\it observed} quasars only would be much smaller at high redshifts than 
shown in Figure 5. 

\subsection{Star-forming Galaxies}

Galaxies with ongoing star-formation are another obvious source of Lyman
continuum photons. 
A composite UV luminosity function of Lyman-break galaxies at 
$z\approx 3$ has been recently derived by Dickinson (1998). It is based
on spectroscopically and photometrically selected 
galaxies from the ground-based and HDF samples, and spans about a factor
50 in luminosity from the faint to the bright end. Because of the uncertanties
that still remain in the rescaling of the HDF data points to the ground-based 
data, the comoving luminosity density at 1500 \AA\ is estimated to vary 
within the range 1.6 to $3.5\times 10^{26}\emunits$. Since the 
rest-frame UV continuum at 1500 \AA\ (redshifted into the visible band for a
source at $z\approx 3$) is dominated by the same short-lived, massive stars
which are responsible for the emission of photons shortward of the Lyman edge,
the needed conversion factor, about one ionizing photon every 10 photons at
1500 \AA, is fairly insensitive to the assumed IMF and is independent of the
galaxy history for $t\gg 10^{7.3}\,$ yr. 

Figure 6 shows the Lyman continuum luminosity function of galaxies 
at $z\approx 3$ [at all ages $\gta 0.1$ Gyr one has $L(1500)/L(912)\approx 
6$ for a Salpeter mass function and constant star formation rate], 
compared to the distribution of QSO luminosities at the same redshift. The 
comoving ionizing emissivity due to Lyman-break galaxies is $4.2\pm 1.5
\times 10^{25}\emunits$, between 2 and 4 times higher than the 
estimated quasar contribution at $z=3$. 
\begin{figure}
\centerline{\epsfig{file=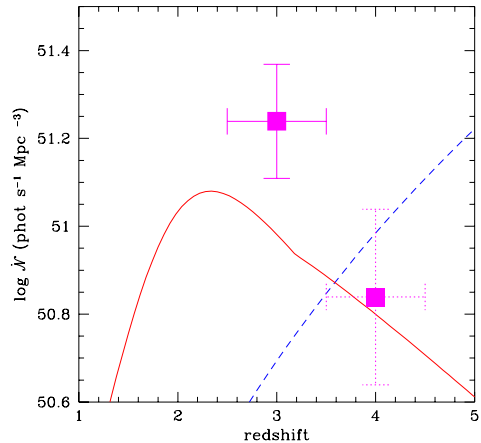,height=7.5cm,width=7.5cm}}
\caption{\em Comoving emission rate of hydrogen Lyman continuum photons ({\it solid
line}) from QSOs, compared with the minimum rate ({\it dashed line}) which is
needed to fully ionize a fast recombining (with clumping factor $C=30$) 
universe with $\Omega_bh_{50}^2=0.08$. Models based on photoionization by 
quasar sources appear to fall short at $z=5$. The data points
with error bars show the estimated contribution of star-forming galaxies 
at $z\approx 3$ and, with significantly larger uncertainties, at $z\approx 4$. 
The fraction 
of Lyman continuum photons which escapes the galaxy \HI layers into the 
intergalactic medium is taken to be $f_{\rm esc}=0.5$.
\label{fig5}} 
\end{figure}
This number neglects any correction for intrinsic \HI absorption: the 
data points plotted in Figure 6 assume a value of $f_{\rm esc}=0.5$ 
for the unknown fraction of ionizing photons which escapes the galaxy \HI 
layers into the intergalactic medium. 

The LF of Lyman-break galaxies at $z\gta 4$ is highly uncertain. An analysis
of the $B$-band dropouts in the HDF -- candidate star-forming objects at
$3.5<z<4.5$ -- seems to imply a decrease in the comoving UV galaxy emissivity
by about a factor of 2.5 in the interval $2.75\lta z\lta 4$ (Madau \etal 1996,
1998c). In this sense star-forming galaxies with SFR in excess of 
$0.5\sfr$ may 
have a negative evolution with lookback time similar to the one observed in 
bright QSOs, but the error bars are still rather large. Adopting a $L(1500)$ 
to $L(912)$ conversion factor of 6, we estimate a comoving ionizing emissivity 
of $1.7\pm 1.1 \times 10^{25} f_{\rm esc} \emunits$ at $z\approx 4$. One 
should note that, while highly reddened sub-mm sources at high redshifts 
(Lilly, this volume) would be missed by 
the dropout color technique (which isolates sources that have blue colors in 
the optical and a sharp drop in the 
rest-frame UV), it seems unlikely that very dusty objects (with $f_{\rm esc}
\ll 1$) would contribute in any significant manner to the ionizing 
metagalactic flux. 

\subsection{Reionization of the Universe}

In inhomogeneous reionization scenarios, the history of the transition from a
neutral IGM to one that is almost fully ionized can be statistically 
described by
the evolution with redshift of the {\it volume filling factor} or porosity
$Q(z)$ of \HII, \HeII, and \HeIII regions. The radiation emitted by spatially
clustered stellar-like and quasar-like sources -- the number densities and 
luminosities of which may change rapidly as a function of redshift -- 
coupled with
absorption processes in a medium with a time-varying clumping factor, all
determine the complex topology of neutral and ionized zones in the universe.
When $Q<<1$ and the radiation sources are randomly distributed, the ionized
regions are spatially isolated, every UV photon is absorbed somewhere in the
IGM, and the ionization process cannot be described as due to a statistically
homogeneous radiation field. As $Q$ grows, the crossing of 
ionization fronts becomes more and more common, and the neutral phase shrinks 
in size until the reionization process is completed at the ``overlap'' epoch, 
when every point in space is exposed to Lyman continuum radiation. 
\begin{figure}
\centerline{\epsfig{file=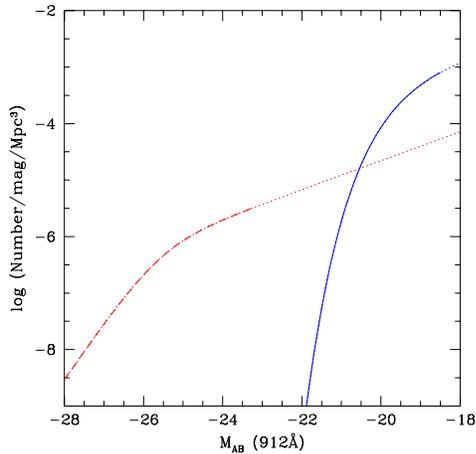,height=7.5cm,width=7.5cm}}
\caption{\em The $912\,$\AA\ luminosity function of galaxies at $z\approx 
3$ ({\it
solid line}), compared to the distribution of QSO luminosities at the same
redshift ({\it dashed line}). 
The former assumes a Salpeter IMF with constant 
constant star formation rate (age=1 Gyr): $M_\AB(912\,$\AA)$=-19$ corresponds 
to a rate of $13\sfr$. The solid and dashed lines represent functional 
fits to the data points, and the dotted lines their extrapolation. 
\label{fig6}} 
\end{figure}
The filling factor of \HII regions in the universe, $Q_\nHII$, is equal at any 
given instant $t$ to the integral
over cosmic time of the rate of ionizing photons emitted per hydrogen atom and
unit cosmological volume by all radiation sources present at earlier epochs, 
$\int_0^t \dot n_{\rm ion}(t')dt'/\bar{n}_\nH(t')$, minus the rate of 
radiative recombinations, $\int_0^t Q_\nHII(t')dt'/\bar{t}_{\rm rec}(t')$
(Madau \etal 1998b). Differentiating one gets
\begin{equation}
{dQ_\nHII\over dt}={\dot n_{\rm ion}\over \bar{n}_\nH}-{Q_\nHII\over 
\bar{t}_{\rm rec}}.  \label{eq:qdot}
\end{equation}
It is this simple differential equation -- and its equivalent for
expanding helium zones -- that statistically describes the transition
from a neutral universe to a fully ionized one, independently, for a given
emissivity, of the complex and possibly short-lived emission histories of
individual radiation sources, e.g., on whether their comoving space density is
constant or actually varies with cosmic time.
Here $\dot n_{\rm ion}$ is the emission rate of ionizing photons per unit proper
volume, $\bar{n}_\nH$ is the mean hydrogen density of the expanding 
IGM, $\bar{n}_\nH(0)=1.7\times 10^{-7}$ $(\Omega_b h_{50}^2/0.08)$ cm$^{-3}$, 
\begin{equation}
\bar{t}_{\rm rec}=[(1+2\chi) \bar{n}_\nH \alpha_B\,C]^{-1}=0.3\, {\rm Gyr} 
\left({1+z\over 4}\right)^{-3} C_{30}^{-1} 
\end{equation}
is the volume-averaged gas recombination timescale, $\alpha_B$ is the 
recombination coefficient to the excited states of hydrogen,
$\chi$ the helium to hydrogen cosmic abundance ratio, $C\equiv \langle
n_\nHII^2\rangle/\bar{n}_\nHII^2$ is the ionized hydrogen clumping factor, and 
a gas temperature of $10^4\,$K has been assumed. Clumps which are dense
and thick enough to be self-shielded from UV radiation will stay neutral and
will not contribute to the recombination rate. 

An empirical determination of 
the  clumpiness of the IGM at high redshifts is hampered by our poor 
knowledge of the ionizing background intensity and the typical size and 
geometry of the absorbers. Numerical N-body/hydrodynamics simulations of 
structure formation in the IGM within the framework of CDM dominated 
cosmologies have 
recently provided a definite picture for the origin of intervening absorption 
systems, one of an interconnected network of sheets and filaments, with 
virialized systems located at their points of intersection.  In the simulations
of Gnedin \& Ostriker (1997), for example, the clumping factor rises above 
unity when the collapsed fraction of baryons becomes non negligible, i.e. 
$z\lta 20$, and grows to $C\gta 10$ (40) at $z\approx 8$ (5) (because of finite 
resolution effects, numerical simulations will actually underestimate 
clumping): the recombination timescale is much shorter than that for a uniform 
IGM, and always shorter than the expansion time.
 
For an IGM with $C\gg 1$, and in the case $\dot n_{\rm ion}$ and $C$ do not 
vary rapidly over a timescale $\bar{t}_{\rm rec}$, one can expand around $t$ 
to find 
\begin{equation}
Q_\nHII(t)\approx {\dot n_{\rm ion}\over \bar{n}_\nH}\bar{t}_{\rm rec}. 
\label{eq:qa}
\end{equation}
The porosity of ionized bubbles is then approximately given by the number
of ionizing photons emitted per hydrogen atom in one recombination time. In 
other words, because of hydrogen recombinations, only a fraction $\bar{t}_{\rm 
rec}/t$
($\sim$ a few per cent at $z=5$) of the photons emitted above 1 ryd is
actually used to ionize new IGM material. The universe is completely reionized
when $Q_\nHII=1$, i.e. when 
\begin{equation}
\dot n_{\rm ion} \bar{t}_{\rm rec}=\bar{n}_\nH. 
\label{eq:qone}
\end{equation}

\section{A POPULATION OF EARLY DWARFS}

As $\bar{t}_{\rm rec} \ll t$ at high redshifts, it is possible to compute 
at any given epoch a critical value for the photon emission rate per unit 
cosmological comoving volume, $\dot {\cal N}_c$,
independently of the (unknown) previous emission history of the universe: only
rates above  this value will provide enough UV photons to ionize the IGM by 
that epoch. One can then compare our determinations of $\dot {\cal N}_c$ to the estimated contribution from QSOs and star-forming galaxies. 
Equation (\ref{eq:qone}) can then be rewritten as
\begin{equation}
\dot {\cal N}_c={\bar{n}_\nH(0)\over \bar{t}_{\rm rec}(z)}=(10^{51.2}\,
\ndotunits)\, C_{30} \left({1+z\over 6}\right)^{3}. 
\label{eq:caln}
\end{equation}
The uncertainty on this critical rate is difficult to estimate, as it depends 
on the clumping factor of the IGM (scaled in the expression above to the 
value inferred at $z=5$ from numerical simulations)
and the nucleosynthesis constrained baryon density. A quick exploration of the 
available parameter space indicates that the uncertainty on $\dot 
{\cal N}_c$ could easily be of order $\pm 0.2$ in the log. The 
evolution of the critical rate as a function of redshift is plotted in Figure 
5. While $\dot {\cal N}_c$ is comparable to the quasar contribution at 
$z\gta 3$, there is some indication of a significant deficit of Lyman 
continuum photons at $z=5$. For bright, massive galaxies to produce enough UV 
radiation at
$z=5$, their space density would have to be comparable to the one observed at
$z\approx 3$, with most ionizing photons being able to escape freely from the
regions of star formation into the IGM. This scenario may be in conflict with  
direct observations of local starbursts below
the Lyman limit showing that at most a few percent of the stellar ionizing
radiation produced by these luminous sources actually escapes into the IGM 
(Leitherer \etal 1995). If, on the other 
hand, faint QSOs with (say) $M_\AB=-19$ at rest-frame ultraviolet frequencies 
were to provide {\it all} the required ionizing flux, their 
comoving space density would be such ($0.0015\,$Mpc$^{-3}$) that about 50 
of them would expected in the HDF down to $I_\AB=27.2$.  At $z\gta 5$, they 
would appear very red in $V-I$ as the \Lya forest is shifted into the visible.
This simple model can be ruled out, however, as there is only a handful (7) 
of sources in the HDF with $(V-I)_\AB>1.5$ mag down to this magnitude limit.
  
It is interesting to convert the derived value of $\dot {\cal N}_c$  
into a ``minimum'' star formation rate per unit (comoving) volume, $\dot 
\rho_*$):
\begin{equation}
{\dot \rho_*}(z)=\dot {\cal N}_c(z) \times 10^{-53.1} f_{\rm esc}^{-1}
\end{equation}
\begin{equation}
\approx 0.013 f_{\rm esc}^{-1} \left({1+z\over 6}\right)^3\ \sfrd 
\label{eq:sfr} 
\end{equation}  
(Salpeter IMF with solar metallicity), comparable with the value directly 
``observed'' (i.e., uncorrected for dust reddening) at $z\approx 3$ 
(Dickinson 1998).  The same massive stars that dominate the
Lyman continuum flux also manufacture and return most of the metals to the 
ISM. In the approximation of instantaneous recycling, the rate of ejection of 
newly sinthesized heavy elements which is required
to keep the universe ionized at redshift $z$ is, from equation (\ref{eq:sfr}),
\begin{equation}
{\dot \rho_Z}(z)=y(1-R){\dot \rho_*}(z) \gta 3.5\times 10^{-4}
 \end{equation}
\begin{equation}
\times \left({y\over 2 Z_\odot}\right) \left({1+z\over 6}\right)^3 f_{\rm esc}^{-1}\, 
\sfrd, \label{eq:mfr}
\end{equation}
where $y$ is the net, IMF-averaged ``yield'' of returned metals, $Z_\odot=0.02$,
and $R\approx 0.3$ is the mass fraction of a generation of stars that is 
returned to the interstellar medium. At $z=5$, and over a timescale of $\Delta 
t=0.5\,$ Gyr (corresponding to a formation redshift $z_f=10$) such a rate would 
generate a mean metallicity per baryon in 
the universe of
\begin{equation}
\langle Z \rangle\approx {8\pi G {\dot \rho_Z}(5) \Delta t\over 3 H_0^2 
\Omega_b}\gta 0.002 \left({y\over 2 Z_\odot}\right) f_{\rm esc}^{-1}Z_\odot,
\end{equation}
comparable with the level of enrichment observed in the \Lya forest at
$z\approx 3$ (Songaila 1997): more than 2\% of the present-day stars would need 
to have formed by $z\sim 5$. It has been recently suggested (Miralda-Escud\'e 
\& Rees 1998) that a large number of low-mass galactic halos, expected to form at early times in hierarchical 
clustering models, might be responsible for photoionizing the IGM at 
these epochs. 
According to spherically-symmetric simulations (Thoul \etal 1996), 
photoionization heating by the UV background flux 
that builds up after the overlapping epoch completely suppresses the cooling 
and collapse of gas inside the shallow potential wells of halos 
with circular velocities $\lta 35\,\kms$. Halos with circular speed 
$v_{\rm circ}=50\,\kms$, corresponding in top-hat spherical collapse to a 
virial temperature $T_{\rm vir}=0.5\mu m_p v_{\rm circ}^2/k\approx 10^{5}\,$K 
and halo mass $M=0.1v_{\rm circ}^3/GH\approx 4\times 10^9 [(1+z)/6]^{-3/2}
h_{50}^{-1}\, \msun$, appear instead largely immune to this external feedback
(but see Navarro \& Steinmetz 1997). In these systems rapid cooling by 
atomic hydrogen can then take place and a significant fraction, 
$f\Omega_b$, of their total mass may be converted into stars over a timescale 
$\Delta t$ comparable to the Hubble time (with $f$ close 
to 1 if the efficiency of forming stars is high). If high-$z$ dwarfs with 
star formation rates $f\Omega_bM/\Delta t\sim 0.3 f_{0.5} \Delta t_{0.5}^{-1}\, 
\sfr$  were actually responsible for keeping the universe ionized at $z\sim 
5$, their comoving space density would have to be 
\begin{equation}
{0.013\, f_{\rm esc}^{-1}\sfrd\over 0.6 f {\, \rm M_\odot yr^{-1}}}
\sim 0.1 \left({f_{\rm esc} f\over 0.25}\right)^{-1} {\,\rm Mpc^{-3}},
\end{equation}
two hundred times larger than the space density of present-day galaxies 
brighter than $L^*(4400)$, and about five hundred times larger than that of 
Lyman-break objects at $z\approx 3$ with $M<M^*_\AB(1500)$,
i.e. with star formation rates in excess of $10\,\sfr$. Only a rather
steep luminosity function, with Schechter slope $\alpha\sim 2$, would
be consistent with such a large space density of faint dwarfs and, at the same
time, with the paucity of brighter $B$- and $V$-band dropouts observed in the
HDF. The number density on the sky would be $\approx 0.2\,$ arcsec$^{-2}$, 
corresponding to more than three thousands sources in the HDF. With a typical 
apparent magnitude at $z=5$ of $I_\AB \sim 29.5$ mag 
(assuming $f=0.5$), these are too faint to be detected by the {\it HST}, 
but within the range of the proposed NGST.
A higher density of sources -- which would therefore have to originate from 
lower amplitude peaks -- would be required if the typical efficiency of star 
formation and/or the escape fraction of ionizing photons were low, $(f, f_{\rm 
esc})\ll 1$. In this case the dwarfs could still be detectable  
if a small fraction of the gas turned into stars in very short bursts.

\section*{ACKNOWLEDGEMENTS}
 
I acknowledge numerous discussions with my collaborators, M. Della Valle, 
F. Haardt, N. Panagia, and M. Rees. Support for this work was provided
by NASA through ATP grant NAG5-4236.

\end{document}